\documentclass{PoS}
\usepackage{url}
\usepackage{parskip}

\usepackage{xspace}

\def\openqcd{{\tt openQCD}\xspace}
\def\simgrid{{\tt SimGrid}\xspace}
\def\ym{{\tt ym1}\xspace}
\def\ymin{{\tt ym1.in}\xspace}
\def\qcdp{{\tt qcd1}\xspace}

\newcommand{\preprintline}{\newline
\vskip -4.0cm
\rightline{\parbox{3.8cm}{\large\rm CERN-TH-2017-027}} 
\vspace{2.4cm}}

\title{Platform independent profiling of a QCD code  \preprintline}

\ShortTitle{Platform independent profiling of a QCD code}

\author{\speaker{Marina Krsti\'c Marinkovi\'c}\\
        CERN, TH Department, 1211 Geneve, Switzerland\\
        School of Mathematics, Trinity College Dublin, Dublin 2, Ireland\\
        E-mail: \email{Marina.Marinkovic@cern.ch}}

\author{Luka Stanisic\\
        Inria Bordeaux Sud Ouest, Bordeaux, France\\
        E-mail: \email{Luka.Stanisic@inria.fr}}

\abstract{The supercomputing platforms available for high performance computing based research evolve at a great rate. However, this rapid development of novel technologies requires constant  adaptations and optimizations of the existing codes for each new machine architecture. In such context, minimizing time of efficiently porting the code on a new platform is of crucial importance. A possible solution for this common challenge is to use simulations of the application that can assist in detecting performance bottlenecks. Due to prohibitive costs of classical cycle-accurate simulators, coarse-grain simulations are more suitable for large parallel and distributed systems. We present a procedure of implementing the profiling for \openqcd code \cite{openQCD} through simulation, which will enable the global reduction of the cost of profiling and optimizing this code commonly used in the lattice QCD community. Our approach is based on well-known \simgrid simulator \cite{SimGrid}, which allows for fast and accurate performance predictions of HPC codes. Additionally, accurate estimations of the program behavior on some future machines, not yet accessible to us, are anticipated.}

\FullConference{34th annual International Symposium on Lattice Field Theory\\
		24-30 July 2016\\
		University of Southampton, UK}

\begin{document}
\section{Introduction}
\label{sec:intro}
The numerical simulations in lattice QCD are large scale. Besides constant algorithmic development, they require massive parallelization and consume a substantial amount of resources of the leading supercomputing centers in Europe and worldwide. In order to efficiently exploit ever-changing HPC architectures, one continually needs to adapt and optimize the existing codes for new hardware.
To develop a lattice QCD code that scales well on diverse platforms, besides the access to different machines, the developer needs to be able to perform large experimental campaigns in a fairly short period of time. 
It would therefore be advantageous to design a universal procedure that would allow for benchmarking of the native code executions on a wide range of supercomputing platforms. 
We propose a way to use an adaptable simulator for parallel systems, \simgrid \cite{SimGrid} in combination with \openqcd \cite{openQCD} code, in order to measure the scalability of the latter without accessing the target machine. The procedure presented here can in the same way be applied to other up-to-date QCD codes.

The aim of this setup is to speed up the benchmarking of the chosen QCD code on an unknown machine and to enable benchmarking of platforms to which the user does not yet have access. 
Moreover, many worlds' supercomputing centers, after awarding computer time, let the choice of the suitable machine within the given center to the users. Being able to estimate the optimal choice of the machine with the appropriate number of nodes in these situations in timely manner and without wasting computational resources is another possible application of this procedure. 

In Section \ref{sec:openQCD}, we briefly describe the program package \openqcd and perform a conventional native set of benchmarks for a pure gauge simulation code. \simgrid simulation framework is introduced in Section \ref{sec:SimGrid}. We outline a procedure for simulating \openqcd programs using \simgrid in Section \ref{sec:profiling}. After reporting our preliminary results, we discus next steps envisioned for this study in Sections \ref{sec:results} and \ref{sec:future}, respectively. Finally, we conclude and address possible future applications of this work in Section \ref{sec:summary}.
\section{\openqcd: benchmarking and scaling}
\label{sec:openQCD}
The simulation package \openqcd contains programs for generating both pure gauge and dynamical QCD configurations, the latter with a choice of {$O(a)$ improved Wilson fermion action}. 
 Although original motivation was to create a portable simulation tool for recently introduced open boundary conditions \cite{Luscher2011a}, the programs in this package allow for several different choices of boundary conditions\footnote{Extending the code to include QED field dynamics and an additional choice of C* boundary conditions \cite{Wiese1992a, Lucini2015a} is in progress.}: open, periodic, Schr\"odinger Functional (SF), and open-SF.
The simulation program is based on the Hybrid Monte Carlo algorithm \cite{Duane1987} and supports parallelization in 0,1,2,3 or 4 dimensions. All the programs in this package are highly optimized for machines with modern Intel or AMD processors, but will run correctly on any system that complies with the ISO C89 and the MPI 1.2 standards. 
Optimized version of the code for BlueGene/Q machines is available from \cite{bgQ}. 
The code is open source with GPL license \cite{openQCD}.

\begin{figure}[bt]
\centering
{{\includegraphics[width=0.3\linewidth,angle=270]{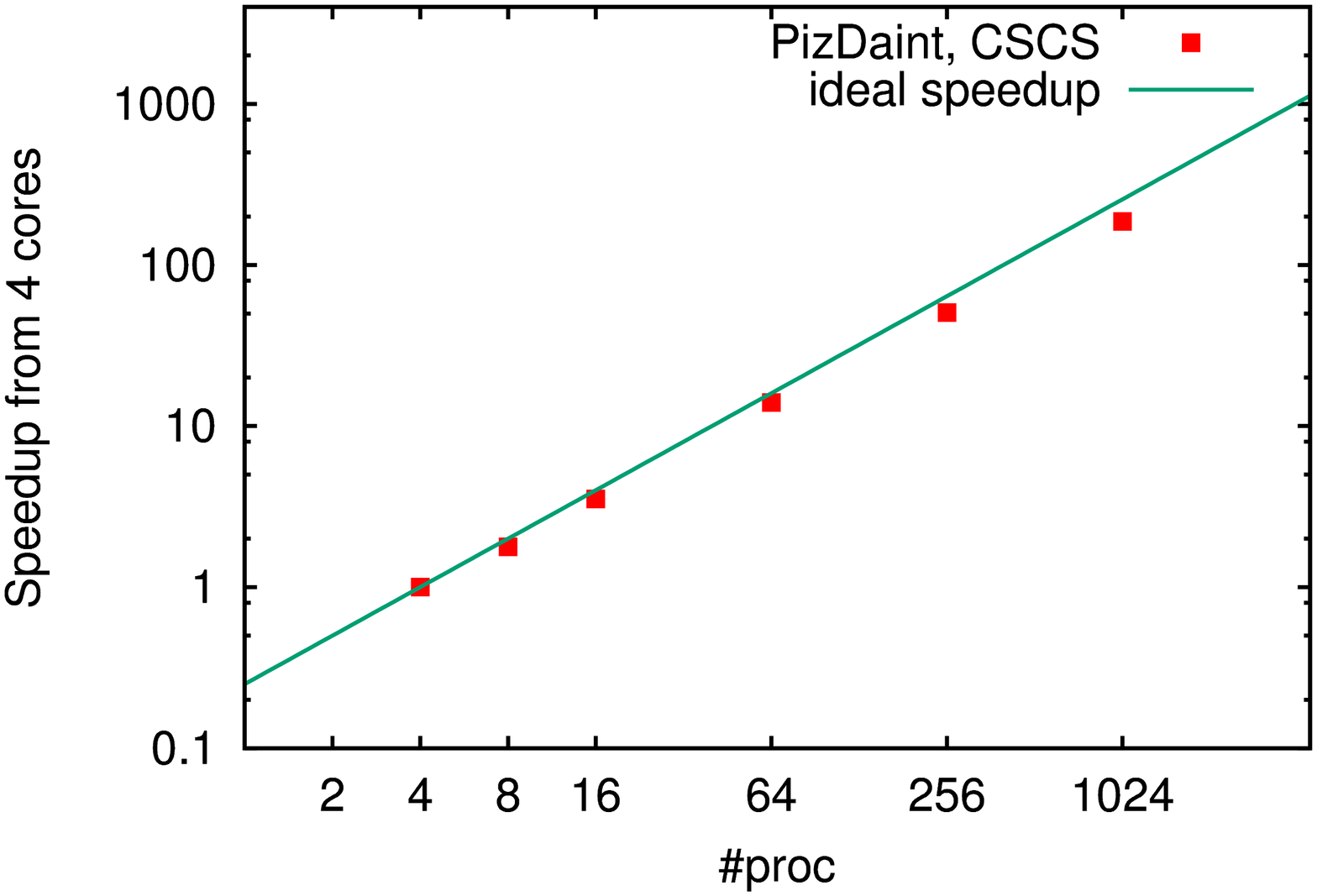}}}
{{\includegraphics[width=0.3\linewidth,angle=270]{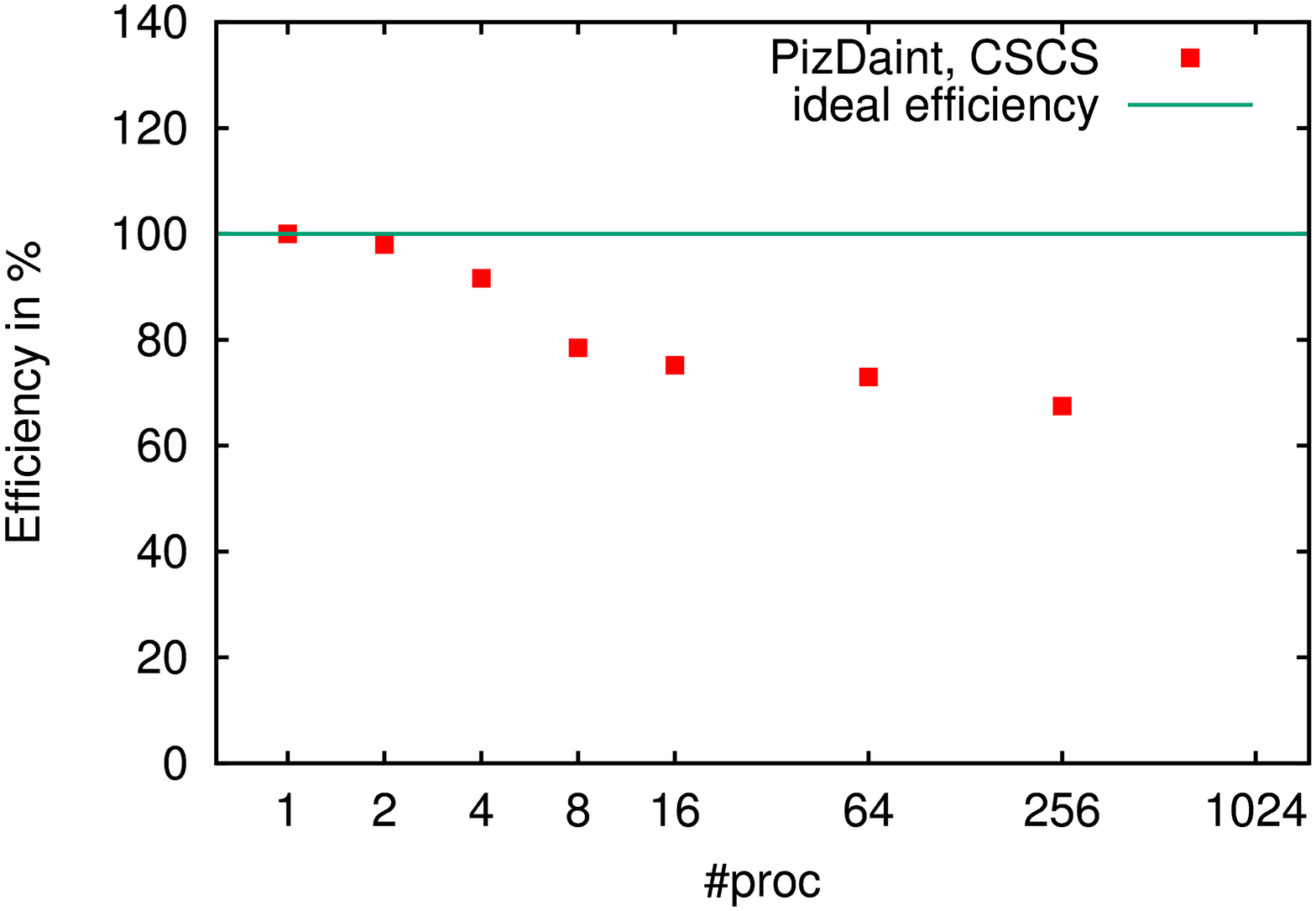}}}
\caption{Strong and weak scaling of the \ym program on a typical Cray machine (PizDaint, CSCS).  Strong scaling plot shown in the left panel is obtained by running \ym on a global lattice of a size $32^4$. Local lattice size used for the weak scaling plot shown in the right panel is $8^4$.  Both plots are obtained with run parameters from the sample input file \ymin distributed with \openqcd \cite{openQCD}.}
\label{fig:scaling}
\end{figure}

We apply the first set of benchmarking on the program \ym that generates ensembles of SU(3) gauge configurations. Exactly which theory is simulated depends on the
parameters passed to the program. For the scaling tests shown in Figure \ref{fig:scaling}, we chose tree-level improved Symanzik gauge action and open-SF boundary conditions.
The scaling tests are performed at the PiZDaint\footnote{As of December 2016, PizDaint machine has been upgraded. The benchmarks performed in this work correspond to the configuration of June 2016.} machine of the CSCS supercomputing center. 
\section{\simgrid simulation framework}
\label{sec:SimGrid}
\simgrid is a versatile simulation framework for distributed systems that allows conducting studies in different fields, such as grid, cloud, peer-to-peer and volunteer computing. In the HPC context, more precisely regarding MPI applications, \simgrid provides SMPI module \cite{pmbs} which can be used in two different modes: offline and online. In the offline mode, a trace of a previous native execution on a supercomputer is replayed in the simulator. Quick and accurate predictions can be obtained, but such results are hard to faithfully extrapolate to different scenarios. The main reason behind this is that modern hardware and software stacks are so complex, that even a minor change to a platform parameter or an application parameter can cause substantial change to the overall executions. On the other hand, online mode, which relies only on the target platform description without any traces, executes directly the application. Therefore, simulations in this mode can mimic much better any modifications to the application or machine setup. Although such approach is much harder to implement and validate, its predictive capabilities are more advanced. There is a plethora of MPI simulators used in HPC community (\cite{SST}, \cite{bigsim}, and many more), but we have chosen \simgrid as its faithful online simulations with accurate modeling of network contention are needed for this type of study.
\begin{figure}[!ht]
\centering
\includegraphics[angle=270,width=\linewidth]{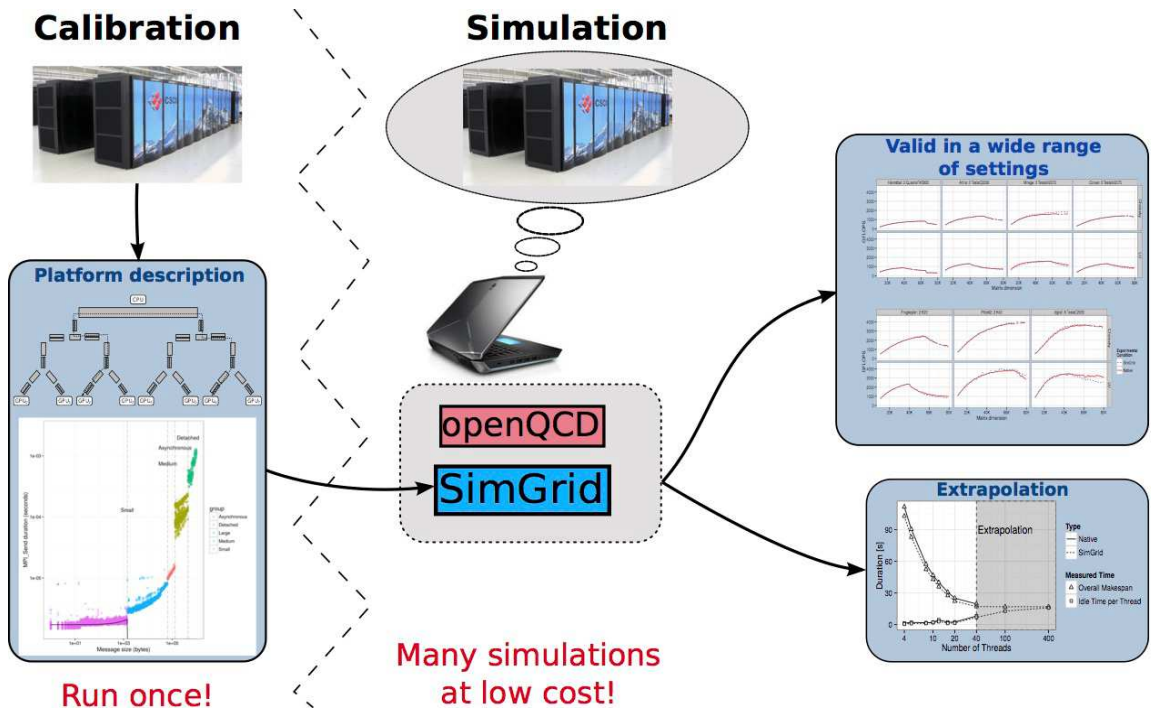}
\caption{Workflow for coupling the \openqcd package and \simgrid. Once platform characteristics have been benchmarked and network models have been instantiated, many simulations can be run at low cost to provide faithful predictions for a wide range of scenarios.}
\label{fig:workflow}
\end{figure}
\section{Methodology for profiling \openqcd with a \simgrid simulator}
\label{sec:profiling}

Workflow for conducting simulations of \openqcd is presented in Figure \ref{fig:workflow}. In the initial phase, target machine needs to be carefully calibrated. One needs to reserve only few nodes of a possibly large supercomputer. These nodes are first benchmarked to capture their characteristics, typically number of processing units and their corresponding speed. Then, a different set of benchmarks containing a wide range of MPI calls is performed to measure the network characteristics and its behavior in the case of contention. The obtained values will be used by \simgrid when instantiating its communication models. These coefficients together with the previously captured platform details are saved in a separate XML file.

Once this platform description file is generated, no more access to the supercomputer is needed. One needs to simply compile \openqcd application with the \simgrid SMPI, instead of classical MPI library. This new program can then be run on a single core of a commodity laptop and provide faithful performance predictions. \openqcd application ``thinks'' that it is executed on a large scale machine using many nodes, while it is actually run inside a single \simgrid process.

Since almost no actual computations and no data transfers are performed, simulation is typically much faster than the native execution and additionally consumes much less memory. This allows for running many profilings at low cost and fully assessing the influence of an application input parameter (e.g. lattice size, bare gauge coupling, etc. 
in the case of a lattice QCD code). Since platform description file can be easily manually edited, one may also explore ``what-if'' scenarios by increasing the number of nodes, number of processing unit, network latency and bandwidths, etc. This can provide invaluable information to the \openqcd developers and users who have a challenging task of correctly dimensioning the problem size they want to study, as well as determine the optimal machine for such workload.

\section{Initial evaluation of \ym}
\label{sec:results}
\begin{figure}[t]
\begin{center}
{\includegraphics[width=0.65\linewidth]{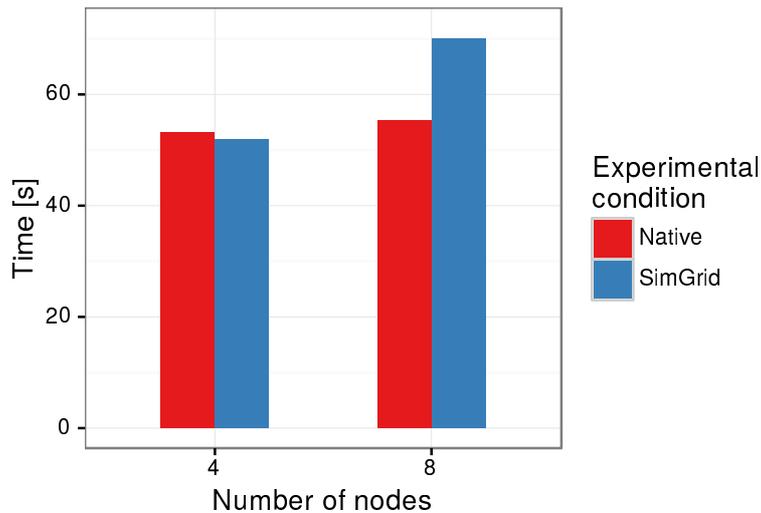}}
\end{center}
\caption{Native (red) and simulated (blue) times on 4 and 8 processors at 2 Dodeca-core Haswell Intel Xeon E5-2680 machine. We simulated four HMC trajectories of pure Yang-Mills simulation program that is a part of \openqcd. Run parameters are taken from the sample input file \ymin distributed with \openqcd \cite{openQCD}.}
\label{fig:performance}
\end{figure}

We used \simgrid to simulate pure Yang-Mills simulation program (\ym) of \openqcd. We compared the time needed to run four trajectories in the generation of pure gauge configurations using Symanzik gauge action and open-SF boundary conditions, on a $8^4$ local lattice. 
The first tests were run on 4 and 8 processors (2 Dodeca-core Haswell Intel Xeon 2,5 GHz 128GB RAM machine) on PlaFRIM computing cluster, corresponding to the global lattice sizes of $16^2\times8^2$ and $16^3\times8$.  Actual execution times of the \ym program were compared to the execution times predicted by \simgrid. The outcome is shown in Figure \ref{fig:performance}. Native \openqcd code demonstrates very good performance, as for these weak scaling experiments the timings between two executions are very similar. Although \simgrid provides an accurate estimation in the case of 4 processors, for 8 nodes prediction is largely overestimating the overall execution duration. 

To understand the source of such discrepancy, we intend to introduce timestamps into \openqcd code. Then, by comparing the traces between native and \simgrid executions, we expect to detect and later correct the error in our models, which will allow for achieving the good estimations for large scale executions. Currently, we suspect that the issues we encountered are caused by bogus network drivers and problems with MPI libraries on PlaFRIM cluster at the time of benchmarking, which lead to inaccurate generation of communication models. However, these assumptions need to be empirically verified.

\section{Future prospects}
\label{sec:future}

For practical applications in lattice QCD, one would need to go beyond testing the program for pure gauge configurations generation \ym. On the example of \openqcd, this would mean to get the good matching of the native and simulated execution times of the \qcdp program. Providing faithful estimates for the dynamical gauge configuration generation program requires accurate predictions of the Dirac matrix inversions. However, since \simgrid has been already successfully used for the simulation of the solvers for sparse linear systems \cite{stanisic:hal-01180272}, we expect a smooth transition to more complex simulation programs such as \qcdp.

Presented study was mostly exploratory, as we wanted to evaluate the feasibility of the simulation approach on \openqcd application and identify the biggest challenges. Once our methodology is extensively validated on a set of small scale experiments and it becomes more robust, we plan to generalize our approach in order to use it for the performance predictions of the large scale arbitrary machines. To accomplish this challenging task, it is mandatory to obtain accurate calibrations of other supercomputers. This is a common prerequisite shared by many researchers in \simgrid community, and thus there is an ongoing work by \simgrid developers in automatizing the calibration process and collecting platform description files for many supercomputers, such as: Stampede and Blue Waters in the USA, MareNostrum in Spain, CSCS in Switzerland, etc.

\section{Summary and Outlook}
\label{sec:summary}
We have proposed a procedure of implementing an intermediate profiling for a QCD code using \simgrid simulator, which is particularly convenient for simulating parallel applications based on MPI programming paradigm. We tested our methodology
on the \ym program of \openqcd, although, the same procedure can be easily applied to other lattice QCD codes.
Our initial prototype works and provides encouraging results, but still needs some improvement in order to reach the desired accuracy, which would allow for doing experiments on a much larger scale.
We expect these to come from the refinement in communication modeling that we are currently pursuing.

In the future, with more robust and automatized solution, we envisage application of the proposed approach on multiple use cases. First, this can help users who are preparing the computer time demands for a machine they yet do not have access to.
Additionally, it may be useful for choosing
the most suitable machine within already awarded computer time allocation, in a timely manner.
Furthermore, this procedure is not only applicable on currently existing machines. It can even be applied on machines that are still in the design phase, once the features such as machine topology, processor speed and communication characteristics become known. Nevertheless, such results would have to be interpreted carefully, as unknown systems could contain some yet undiscovered (and thus not modeled) phenomena.

Finally, source code and raw data generated during this study are publicly available\footnote{\url{http://gitlab.inria.fr/stanisic/openqcd-simgrid/}} to anyone interested in learning more details about this work.

\subsection*{Acknowledgments}

 This work was supported by a grant from the Swiss National Supercomputing Centre (CSCS) under project ID s642. Some experiments presented in this paper were carried out using the PlaFRIM experimental testbed, being developed under the Inria PlaFRIM development action with support from LABRI and IMB and other entities: Conseil Regional d'Aquitaine, FeDER, Universite de Bordeaux and CNRS (see https://plafrim.bordeaux.inria.fr/).
 
\bibliographystyle{unsrt}

\end{document}